\documentclass[conference]{IEEEtran}
\IEEEoverridecommandlockouts
\usepackage{cite}
\usepackage{amsmath,amssymb,amsfonts}
\usepackage{algorithmic}
\usepackage{graphicx}
\usepackage{textcomp}
\usepackage{xcolor}

\usepackage{multirow}

\def\BibTeX{{\rm B\kern-.05em{\sc i\kern-.025em b}\kern-.08em
    T\kern-.1667em\lower.7ex\hbox{E}\kern-.125emX}}
\begin{document}

\title{ViT-DeiT: An Ensemble Model for Breast Cancer Histopathological Images Classification\\

}

\author{\IEEEauthorblockN{ Amira Alotaibi}
\IEEEauthorblockA{\textit{Computer Science Department} \\
\textit{Umm Al-Qura University }\\
Makkah, Saudi Arabia \\
s44280250@st.uqu.edu.sa}
\and
\IEEEauthorblockN{Tarik Alafif}
\IEEEauthorblockA{\textit{Computer Science Department} \\
\textit{Umm Al-Qura University}\\
Makkah, Saudi Arabia  \\
tkafif@uqu.edu.sa}
\and
\IEEEauthorblockN{Faris Alkhilaiwi }
\IEEEauthorblockA{\textit{Natural Products and Alternative Medicine Department} \\
\textit{King Abdulaziz University}\\
Jeddah, Saudi Arabia \\
Faalkhilaiwi@kau.edu.sa}
\and
\IEEEauthorblockN{Yasser Alatawi}
\IEEEauthorblockA{\textit{Pharmacy Practice Department} \\
\textit{University of Tabuk}\\
Tabuk, Saudi Arabia \\
Yasser@ut.edu.sa}
\and
\IEEEauthorblockN{Hassan Althobaiti}
\IEEEauthorblockA{\textit{Computer Science Department} \\
\textit{Umm Al-Qura University}\\
Makkah, Saudi Arabia  \\
hmthobaiti@uqu.edu.sa}

\and
\IEEEauthorblockN{Abdulmajeed Alrefaei}
\IEEEauthorblockA{\textit{Biology Department} \\
\textit{Umm Al-Qura University}\\
Makkah, Saudi Arabia  \\
afrefaei@uqu.edu.sa}

\and
\IEEEauthorblockN{Yousef M Hawsawi}
\IEEEauthorblockA{\textit{Research Center} \\
\textit{King Faisal Specialist Hospital and Research Center}\\
Jeddah, Saudi Arabia \\
hyousef@kfshrc.edu.sa}
\and 

\IEEEauthorblockN{Tin Nguyen}
\IEEEauthorblockA{\textit{Computer Science and Engineering Department} \\
\textit{University of Nevada}\\
Nevada, United State \\
tinn@unr.edu}

}

\maketitle

\begin{abstract}

Breast cancer is the most common cancer in the world and the second most common type of cancer that causes death in women. The timely and accurate diagnosis of breast cancer using histopathological images is crucial for patient care and treatment. Pathologists can make more accurate diagnoses with the help of a novel approach based on image processing. This approach is an ensemble model of two types of pre-trained vision transformer models, namely, Vision Transformer (ViT) and Data-Efficient Image Transformer (DeiT). The proposed ViT–DeiT model classifies breast cancer histopathology images into eight classes, four of which are categorized as benign, whereas the others are categorized as malignant. The BreakHis public dataset was used to evaluate the proposed model. The experimental results showed 98.17\% accuracy, 98.18\% precision, 98.08\% recall, and a 98.12\% F1 score.

\end{abstract}


\section{Introduction}
According to the World Health Organization, 2.3 million women worldwide were diagnosed with breast cancer in 2020, with 685,000 deaths, making it the most prevalent cancer globally. As many as 7.8 million women were diagnosed with breast cancer between 2015 and 2020 \cite{Breastcancer}.

The timely diagnosis of breast cancer can increase the survival rate; hence, several techniques, such as mammography, magnetic resonance imaging, ultrasound, computed tomography, positron emission tomography, biopsy, and microwave imaging, are used in clinics to diagnose this disease \cite{wang2017early}.
Although years of experience are usually required for radiologists to correctly diagnose malignant tumors from histopathological images, experts occasionally differ in their conclusions. The usage of computer-aided diagnosis (CAD) for image diagnosis can help medical experts produce precise decisions \cite{jiang2019breast}.

The Food and Drug Administration  approved the first viable CAD system for second-opinion screening mammography in 1998  \cite{chan2019cad}.
Histopathological images of breast cancer can be employed  for clinical applications to automatically and accurately detect malignant tumors. Moreover, deep learning    algorithms have been extensively employed to improve detection performance. Deep learning     algorithms have successfully escalated  the performance of classification of histopathological images.

Convolutional Neural Networks (CNNs) are extensively implemented in computer vision applications, including the detection of histopathological images. Many studies have been conducted to improve the performance of CNNs for breast cancer image classification \cite{gupta2020analysis,albashish2021deep,kassani2019classification}. The training of Vision Transformer (ViT) with sufficiently large data has been shown to achieve remarkable results.  ViT outperforms comparable state-of-the-art CNNs, with four times less computational effort. Nevertheless, transformers were originally innovated  for natural language processing \cite{vaswani2017attention}. In a transformer,  a sequence of tokens is passed as input, but in ViT \cite{dosovitskiy2020image}, image patches are passed as inputs. 

In this study , an ensemble model based on ViT and Data-Efficient Image Transformer (DeiT) models is proposed for breast cancer histopathological image classification. The main contributions of our work are summarized as follows:
\begin{itemize}
    \item An image classification and design of a new CAD system based on a ViT– DeiT ensemble model is proposed. In this method, the ViT and DeiT models are trained by transfer learning for multiclass classification. 
    \item The proposed ensemble model is used to increase classification reliability and minimize false negatives.
    \item Investigate image magnification dependent and independent approaches on a BreakHis dataset for multi-class classification.
    \item The classification performance was compared with similar studies. In terms of classification accuracy, the ViT–DeiT model outperforms other models.

\end{itemize}

The remaining parts of the paper are organized as follows. We present related works in Section \ref{sec:RelatedWork}, and review dataset in Section \ref{sec:dataset}. We focus on ViT and DeiT in the preliminaries Section \ref{sec:Preliminaries} and introduce our proposed method in Section \ref{sec:METHODOLOGY}. Section \ref{sec:EXPERIMENTS} provides experiments setup, result of proposed model and comparison with similar studies. Section \ref{sec:con} concludes the paper.

\section{Related Work}
\label{sec:RelatedWork}

Recently, studies in the field of breast cancer classification have focused on ultrasound image classification \cite{li2021research,gheflati2022vision,lazo2020comparison,tehrani2020pilot}, biopsy data (CSV  file)  classification  \cite{punitha2021automated,obaid2018evaluating,alagoz2018university}, and histopathological image classification. The latter is the focus of this study, considering that previous studies have not reached sufficient accuracy. 

Several studies have used CNNs to classify breast cancer using public datasets, with satisfactory results  \cite{yari2020deep,gupta2020analysis,albashish2021deep}. Parvin et al. \cite{parvin2020comparative} compared the performance of five CNN architectures. The models were evaluated via magnification-dependent classification using a public dataset named BreakHis. The best results were achieved using inception-v1, showing accuracies of 89\% to 94\% for binary classification.  These results are considered good, as there were few images to train from scratch. Likewise, Agarwal et al. \cite{agarwal2022breast} proposed and analyzed the performance of four CNN-based architectures —VGG-16, VGG-19, MobileNet, and ResNet-50— on the BreakHis dataset. VGG-16 achieved the highest accuracy of 94.67\% for binary classification.

In addition to CNN models, classical networks, such as Xception, have been proposed for image classification and have shown remarkable results.
Sharma et al. \cite{sharma2022xception} used transfer learning for an Xception CNN pre-trained model as a feature extractor and used a support vector machine (SVM)  as a classifier to classify  histopathological images of breast cancer in the BreakHis  dataset. The model achieved accuracies from 94.11\% to 96.25\% for binary classification.

Another deep learning approach proposed by Zhou et al. \cite{zhou2022breast} was based on a resolution adaptive network (RANet) model and anomaly detection with an SVM (ADSVM) for binary and multiclass classification for each magnification factor of the BreakHis dataset. The RANet–ADSVM model was trained and compared with and without a balanced dataset. In the experiments, the RANet–ADSVM approach achieved the highest accuracy of 93.35\% to 99.14\% for multi-class classification with balanced data. Although the RANet–ADSVM method has achieved better performance with a balanced dataset than with an imbalance dataset, there were marked improvements in classification performance.

Seo et al. \cite{seo2022scaling} proposed a method based on the Primal-Dual Multi-Instance SVM model. The method was evaluated for the binary classification of images with magnifications dependent on the BreakHis database. The accuracy ranged from 85.3\% to 89.8\%. Although many studies have used the BreakHis dataset, a large number of them worked on binary classification, and those that worked on multiple classifications did not reach significant results. In addition, very few studies have been conducted on multi-class magnification-independent classification.

\section{BreakHis Dataset}
\label{sec:dataset}
\begin{figure}[htpb]
	\centering
		\includegraphics[width=0.5\textwidth]{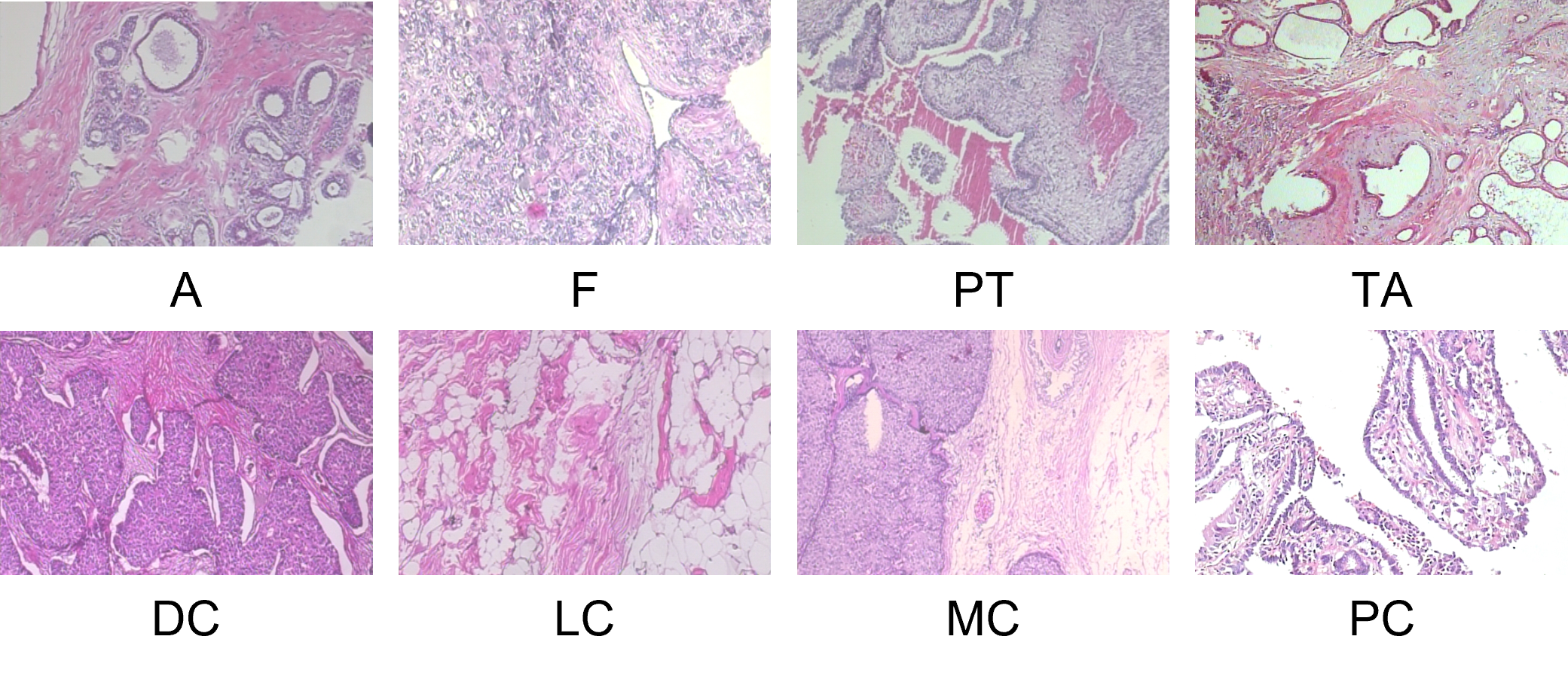}
	\caption{Samples of BreakHis dataset images from different classes at 40× magnification.}
	\label{fig:multi-class}
\end{figure}

BreakHis \cite{spanhol2015dataset} contains microscopic breast tumor biopsy images. The tumors include either benign or malignant tumors. The dataset covers 7,909 images gathered from 82 patients using four magnification factors (40×, 100×, 200×,  and 400×). The dataset contained 2,480 benign tumor images and 5,429 malignant tumor images. The benign tumor images are divided into adenosis (A), tubular adenoma (TA), fibroadenoma (F), and phyllodes tumor (PT). The malignant tumor images are divided into ductal carcinoma (DC), lobular carcinoma (LC), papillary carcinoma (PC), and mucinous carcinoma (MC). The statistics of the BreakHis dataset are shown in  Table \ref{tab:BreakHis}. Fig. \ref{fig:multi-class} shows images of some of the samples at 40× magnification.

\begin{table}[htbp]
\centering
\caption{\label{tab:BreakHis}Number of images for benign and malignant categories in detail.}
\setlength{\tabcolsep}{4.5pt}
\centering
\begin{tabular}{|ll|llll|llll|l|}
\hline
\multicolumn{2}{|l|}{Main category}                                  & \multicolumn{4}{c|}{Benign}                                                            & \multicolumn{4}{c|}{Malignant}                                                         & \multirow{2}{*}{Total} \\ \cline{1-10}
\multicolumn{2}{|l|}{Sub category}                                   & \multicolumn{1}{l|}{A}   & \multicolumn{1}{l|}{F}     & \multicolumn{1}{l|}{TA}  & PT  & \multicolumn{1}{l|}{DC}    & \multicolumn{1}{l|}{LC}  & \multicolumn{1}{l|}{MC}  & PC  &                        \\ \hline
\multicolumn{1}{|l|}{\multirow{4}{*}{\rotatebox[origin=c]{90}{Magnification }\rotatebox[origin=c]{90}{factor}}} & 40X  & \multicolumn{1}{l|}{114} & \multicolumn{1}{l|}{253}   & \multicolumn{1}{l|}{109} & 149 & \multicolumn{1}{l|}{864}   & \multicolumn{1}{l|}{156} & \multicolumn{1}{l|}{205} & 145 & 1,995                  \\[3.5pt] \cline{2-11} 
\multicolumn{1}{|l|}{}                                        & 100X & \multicolumn{1}{l|}{113} & \multicolumn{1}{l|}{260}   & \multicolumn{1}{l|}{121} & 150 & \multicolumn{1}{l|}{903}   & \multicolumn{1}{l|}{170} & \multicolumn{1}{l|}{222} & 142 & 2,081                  \\[3.5pt] \cline{2-11} 
\multicolumn{1}{|l|}{}                                        & 200X & \multicolumn{1}{l|}{111} & \multicolumn{1}{l|}{264}   & \multicolumn{1}{l|}{108} & 140 & \multicolumn{1}{l|}{896}   & \multicolumn{1}{l|}{163} & \multicolumn{1}{l|}{196} & 135 & 2,013                  \\[3.5pt] \cline{2-11} 
\multicolumn{1}{|l|}{}                                        & 400X & \multicolumn{1}{l|}{106} & \multicolumn{1}{l|}{237}   & \multicolumn{1}{l|}{115} & 130 & \multicolumn{1}{l|}{788}   & \multicolumn{1}{l|}{137} & \multicolumn{1}{l|}{169} & 138 & 1,820                  \\[3.5pt] \hline
\multicolumn{2}{|c|}{ Total}                                          & \multicolumn{1}{l|}{444} & \multicolumn{1}{l|}{1,014} & \multicolumn{1}{l|}{453} & 569 & \multicolumn{1}{l|}{3,451} & \multicolumn{1}{l|}{626} & \multicolumn{1}{l|}{792} & 560 & 7,909                  \\[3.5pt] \hline
\end{tabular}

\end{table}

\section{Preliminaries}
\label{sec:Preliminaries}
\subsection{ViT Model}
\label{sec:ViT}
\begin{figure}[htbp]
	\centering
		\includegraphics[width=0.13\textwidth]{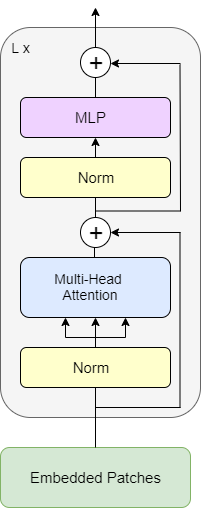}
	\caption{Transformer encoder.}
	\label{fig:Encoder}
\end{figure}

\begin{figure}[htbp]
	\centering
		\includegraphics[width=0.5\textwidth]{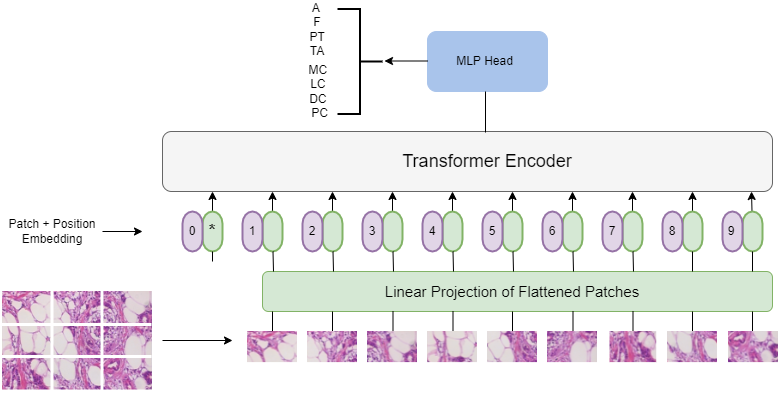}
	\caption{Vision Transformer.}
	\label{fig:ViT}
\end{figure}

The transformer is  widely used for NLP \cite{vaswani2017attention}. The structure of a transformer model comprises an encoder and a decoder. The decoder is not needed in ViT \cite{dosovitskiy2020image}. Therefore, the ViT structure consists of only the encoder for image processing, as shown in Fig. \ref{fig:ViT}. The encoder component consists of normalization layers, a multi-head attention layer, and a feed-forward  layer. The components of the transformer–encoder structure are shown in Fig. \ref{fig:Encoder}. The multi-head attention is a type of self-attention that functions to pay attention to information from various aspects. 

In the ViT model, the image passes through a linear embedding layer before being fed to the encoder. The embedding layer divides the image into equal-sized patches that are flattened into a one-dimensional vector. The position of the embedding is added to the flattened patches, and the class of the embedded image is added. After the encoder processes these inputs, it produces the output. The output passes through the MLP head structure, which performs the classification task. The class is the output of the MLP head structure. The MLP head consists of two connected layers with a GELU activation function \cite{dosovitskiy2020image}. Google AI released more than one model trained on various datasets of different sizes.

\subsection{DeiT Model }
\label{sec:DeiT}
\begin{figure}[htbp]
	\centering
		\includegraphics[width=0.3\textwidth]{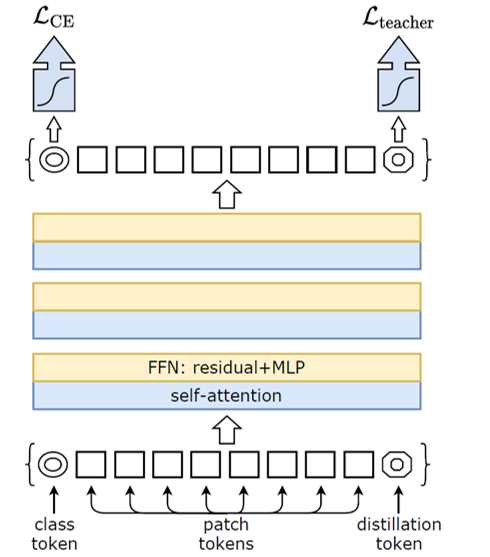}
	\caption{DeiT \cite{touvron2021training}.}
	\label{fig:DeiT}
\end{figure}

Facebook AI introduced DeiT \cite{touvron2021training}. They trained the DeiT model in fewer days and in one machine compared with ViT \cite{dosovitskiy2020image}. The DeiT model was trained on the ImagNet dataset. The model showed improvements over previous ViT models. The structure of DeiT was built based on the ViT model  \cite{dosovitskiy2020image}. They are added with a feed-forward  network (FFN) above the Multi-head self-attention (MSA) layer, which comprises two linear layers separated by GELU activation. As shown in Fig. \ref{fig:DeiT} there is an extra input called a distillation token. This token allows the model to learn from the teacher’s output. The authors attempted soft distillation and hard distillation, with the latter achieving the best results. 

All released checkpoints were pre-trained and fine-tuned on ImageNet-1k only, and no external data were used. This is in contrast with the original ViT model, which used external data, such as JFT-300M dataset and Imagenet-21k, for pre-training.

\section{Proposed Method}
\label{sec:METHODOLOGY}
The proposed method aims to classify histopathological images of breast cancer into eight categories based on magnification-dependent and magnification-independent approaches. As shown in Fig.\ref{fig:VN} the proposed method has three steps, as follows: 
\begin{enumerate}
    \item Preprocess the dataset.
    \item Train the ViT and DeiT models.
    \item Develop the ViT–DeiT ensemble model. 
\end{enumerate}
\subsection{Preprocess the Dataset}	
In BreakHis, the number of images in subcategories is uneven. However, the DC category has the largest number of images (3,451) at different magnifications. The F category has the second largest number of images (1,014), and the other categories has a similar number of images. Thus, imbalanced data lead to overfitting \cite{li2020analyzing}. To avoid the overfitting issue, we propose an undersampling technique.  The undersampling technique reduces the number of samples in a category with a large sample size . Moreover, the undersampling technique is used to balance the dataset before training a model for breast cancer classification at magnification-dependent and -independent approaches. 

\subsection{Train the ViT and DeiT Models}  
To train the models, deep learning requires numerous samples, which transfers learning solves. With a few images, a pre-trained model derived from training on large datasets can be trained, thereby greatly reducing training time. In addition to improving the convergence speed and generalization ability of the model, transfer learning reduces the risk of overfitting. In
order to the number of images in BreakHis is few to train
from scratch, the transfer learning is used. In this study, the transfer learning method was used to train both the ViT model and the DeiT model. However, the models are fine-tuned by placing a prediction head on top of the final hidden state of the class token to classify eight classes. In addition, the ViT and DeiT models were pre-trained on ImageNet with a size of 224$\times$224 and a batch size of 16, which are representations of the base models. Although the models consisted of 12 self-attention heads, the outputs of the heads were combined to produce a final attention score. The final attention score was used to pay attention  to a region of interest. The region of interest represents the cancer cells used for breast cancer classification. However, the attention scores of ViT and DeiT differ because of differences in their structures.

\subsection{Develop the ViT-DeiT Ensemble Model}
\begin{figure}[htbp]
	\centering
		\includegraphics[width=0.45\textwidth]{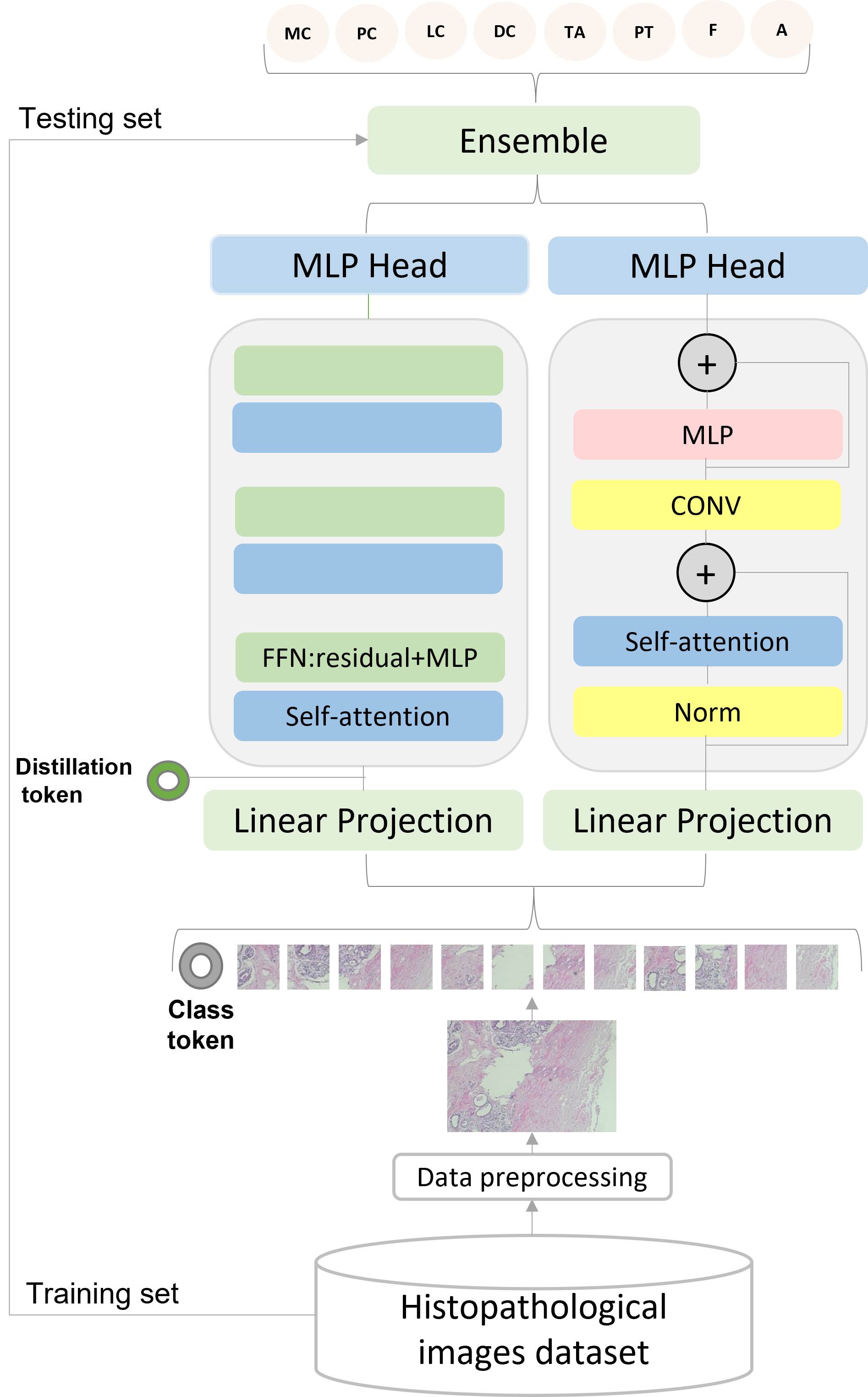}
	\caption{Structure of the proposed ViT–DeiT model.}
	\label{fig:VN}
\end{figure}

The proposed ensemble model uses multi-learning to achieve better performance than that achieved by any single model. Fig. \ref{fig:VN} shows the ViT–DeiT ensemble model. The proposed ensemble model combines the ViT model and the DeiT model, which are two different models. Compared to ViT, the DeiT model uses the distillation token to learn effectively from a teacher. The distillation token is learned through back propagation—that is, by interacting with class and patch tokens through the self-attention layers.

The ensemble technique used in the proposed model is soft voting. The soft voting technique works by assigning the high average probability as the predicted label for each sample. When an image is taken as an input, all models provide a probability values for each class. The probabilities are summed for each class and then divided by the number of classifiers.  The highest probability was assigned for the predicted label, as shown in
 \eqref{eq:softVot}.

\begin{equation}
    \hat{y}=\underset{i}{\operatorname{argmax}}\left\{\frac{1}{N}\sum_{j=1}^n  p_{i j}\right\}
    \label{eq:softVot}
\end{equation}

where $N$ is the number of classifiers, and $p_{ij}$ is the probability of $j^{th}$ classifier for the $i^{th}$ category.

\section{Experiments and Results}

\label{sec:EXPERIMENTS}
\subsection{Experimental Setup}
In the training stage, the value assigned to the learning rate was 1e-4, the weight  decay was 0.001, the batch size was 16, and the number of epochs was 15. These values were selected based on several values tested until the best results were achieved.  The input images were balanced to avoid bias and overfitting in the classification. The images were then divided into 80\% for training and 20\% for testing.

\subsection{Results}
The proposed model was evaluated in a magnification-independent approach, with the same parameters used across the magnification groups. 

As shown in Table \ref{tab:BreakHis}, the BreakHis dataset consists of images grouped into eight categories within two classes, benign and malignant. Accordingly, we evaluated the performance of the breast cancer classification approach for binary classes and multiple classes . To evaluate the effect of the balanced data scenario, we conduct an analysis with and without magnification factors for the classification system based on the ViT–DeiT ensemble model.

In binary classification, the ViT and DeiT covered all images for magnification-independent classification . In multi-class classification, Table \ref{tab:ViTandDeit} shows the accuracy of the ViT and the DeiT models on the test set. The accuracy shows that the performance of ViT and DeiT was close, in which the accuracy of the ViT model was 0.28\% higher than that of the DeiT model. However, the performance of the classification for each category was different, as clearly shown in the TA and F categories in  Table \ref{tab:ModelsR}. After being combined  into the ViT–DeiT ensemble model, the accuracy reached 98.17\% and the performance was further improved, as shown in Table \ref{tab:multiBal}.

\begin{table}[htbp]
\caption{\label{tab:ViTandDeit}Performance of ViT and DeiT before combination.}
\centering
\begin{tabular}{|l|l|l|l|l|}
\hline
Model & Accuracy \% & Precision \% & Recall \% & F1 score \% \\\hline
ViT   & 97.75       & 97.78        & 97.67     & 97.71        \\\hline
DeiT  & 97.47       & 97.47        & 97.41     & 97.43         \\\hline
\end{tabular}
\end{table}

\begin{table}[htbp]
\centering
\caption{\label{tab:ModelsR}Results of the ViT and DeiT models for multiclass magnification-independent classification  in detail.}
\setlength{\tabcolsep}{4pt}
\begin{tabular}{|l|ll|ll|ll|ll|}
\hline
\multirow{2}{*}{\begin{tabular}[c]{@{}l@{}}Sub \\ category\end{tabular}} & \multicolumn{2}{l|}{Accuracy \%}   & \multicolumn{2}{l|}{Precision \%}  & \multicolumn{2}{l|}{Recall \%}     & \multicolumn{2}{l|}{F1-score \%}   \\ \cline{2-9} 
                                                                         & \multicolumn{1}{l|}{ViT}   & DeiT  & \multicolumn{1}{l|}{ViT}   & DeiT  & \multicolumn{1}{l|}{ViT}   & DeiT  & \multicolumn{1}{l|}{ViT}   & DeiT  \\ \hline
A                                                                        & \multicolumn{1}{l|}{99.72} & \textbf{99.72} & \multicolumn{1}{l|}{98.11} & 99.04 & \multicolumn{1}{l|}{\textbf{100}}   & \textbf{99.04} & \multicolumn{1}{l|}{99.05} & \textbf{99.04} \\ \hline
DC                                                                       & \multicolumn{1}{l|}{99.58} & 99.44 & \multicolumn{1}{l|}{\textbf{100}}   & \textbf{100}   & \multicolumn{1}{l|}{96.63} & 95.51 & \multicolumn{1}{l|}{98.29} & 97.70 \\ \hline
F                                                                        & \multicolumn{1}{l|}{98.73} & 99.16 & \multicolumn{1}{l|}{96.55} & 96.67 & \multicolumn{1}{l|}{93.33} & 96.67 & \multicolumn{1}{l|}{94.92} & 96.67 \\ \hline
LC                                                                       & \multicolumn{1}{l|}{99.16} & 99.02 & \multicolumn{1}{l|}{94.79} & 94.74 & \multicolumn{1}{l|}{98.91} & 97.83 & \multicolumn{1}{l|}{96.81} & 96.26 \\ \hline
MC                                                                       & \multicolumn{1}{l|}{99.86} & 99.58 & \multicolumn{1}{l|}{\textbf{100}}   & 98.82 & \multicolumn{1}{l|}{98.84} & 97.67 & \multicolumn{1}{l|}{99.42} & 98.25 \\ \hline
PC                                                                       & \multicolumn{1}{l|}{99.72} & 99.44 & \multicolumn{1}{l|}{97.94} & 96.91 & \multicolumn{1}{l|}{\textbf{100}}   & 98.95 & \multicolumn{1}{l|}{98.96} & 97.92 \\ \hline
PT                                                                       & \multicolumn{1}{l|}{98.73} & 98.87 & \multicolumn{1}{l|}{94.87} & 94.94 & \multicolumn{1}{l|}{93.67} & 94.94 & \multicolumn{1}{l|}{94.27} & 94.94 \\ \hline
TA                                                                       & \multicolumn{1}{l|}{\textbf{100}}   & \textbf{99.72} & \multicolumn{1}{l|}{\textbf{100}}   & 98.68 & \multicolumn{1}{l|}{\textbf{100}}   & 98.68 & \multicolumn{1}{l|}{\textbf{100}}   & 98.68 \\ \hline
\end{tabular}

\end{table}

\begin{figure}[htpb]
	\centering
		\includegraphics[width=0.5\textwidth]{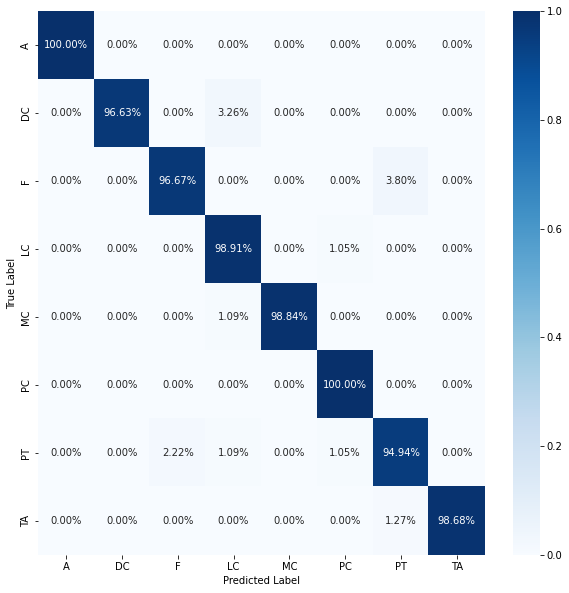}
	\caption{Confusion matrix of ensemble model.}
	\label{fig:conf}
\end{figure}

Accuracy can be used to judge a model’s classification ability, but it cannot reflect specific details. When the classification model makes predictions, the confusion matrix indicates the prediction details of each category by comparing the predicted result with the actual value. As shown in Fig. \ref{fig:conf} the confusion matrix was used to further evaluate the classification ability of the ViT–DeiT model.

In addition, Table \ref{tab:multiBal} shows other metrics, such as precision, recall, and F1 scores, for multi-class classification. As shown in Table \ref{tab:multiBal}, the performance of the models was affected by the data balance method. In the magnification-independent evaluation, using the balanced data scenario, the multi-class analysis classification accuracy was approximately 3.99\% higher than without data balancing. In addition, other metrics demonstrated that the data balance method improved classification performance. As a result of the data balance method, the BreakHis dataset was unevenly distributed, which could explain this finding. 

\begin{table}[htbp]
\caption{\label{tab:multiBal}Performance of the ViT–DeiT model (\%) with a magnification-independent multi-class classification task with a balanced and an imbalanced dataset.}
\centering
\begin{tabular}{|l|l|l|l|l|}
\hline
Method            & Accuracy  & Precision  & Recall  & F1 score  \\ \hline
Imbalance dataset & 94.18       & 94.62        & 93.08     & 93.80        \\\hline
Balance dataset   & \textbf{98.17}       & \textbf{98.18}        & \textbf{98.08}     & \textbf{98.12}        \\ \hline
\end{tabular}
\end{table}

\begin{figure}[htpb]
	\centering
		\includegraphics[width=0.5\textwidth]{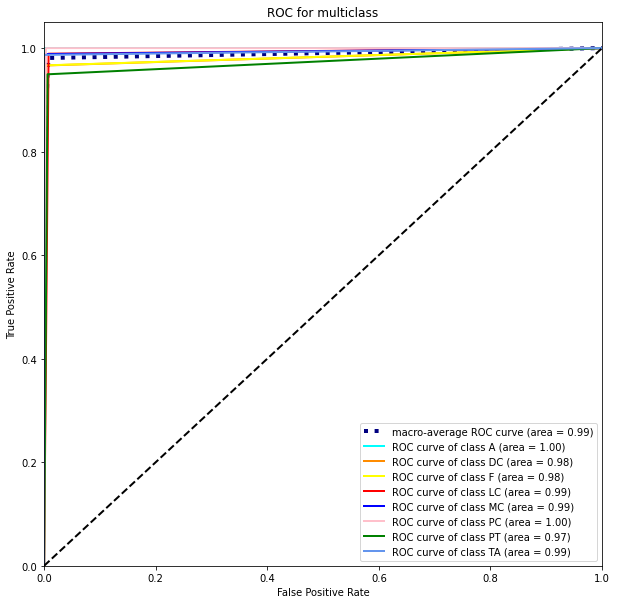}
	\caption{ROC curve illustrating the classification ability of ViT–Net.}
	\label{fig:ROC}
\end{figure}

In addition, a receiver operating characteristic (ROC) curve was constructed to evaluate the performance of the ViT–DeiT model. Fig. \ref{fig:ROC} shows the ROC curve of images attained at the magnification-dependent evaluation for multi-class classification. Moreover, the worst classification performance was observed in the PT ROC curve, causing the proposed model to be confused between PT and the other categories.  In addition, Fig. \ref{fig:ROC} shows the area under the curve (AUC). The ViT–DeiT model achieved an AUC of 0.99, whereas A and PC obtained the best AUC of 1.00.

In general, for magnification-independent evaluation, an accuracy of 98.17\% was obtained for the multi-class classification. However, for magnification-dependent evaluation, the best accuracy of 99.43\% for multi-class classification was obtained at 40× magnification. The best precision of 99.38\%, the best recall of 99.46\%, and the best F1 score of 99.40\% were achieved at the same magnification, as shown in  Table~\ref{tab:mag}. In addition, the performance of the classification obtained at 100× and 200× were close, indicating that  the amount of magnification was convergent.

\begin{table}[htbp]
\caption{\label{tab:mag}Performance of the proposed ViT-DeiT model (\%) for all magnification factors for the multi-class classification task.}
\centering
\begin{tabular}{|l|l|l|l|l|}
\hline
Magnification & Accuracy & Precision & Recall & F1 score \\ \hline
40X           & \textbf{99.43}    & \textbf{99.38}     & \textbf{99.46}  & \textbf{99.40}     \\ \hline
100X          & 98.34    & 98.31     & 98.51  & 98.35     \\\hline
200X          & 98.27    & 98.32     & 98.27  & 98.23     \\\hline
400X          & 98.82    & 98.57     & 98.78  & 98.65    \\ \hline
\end{tabular}
\end{table}

\begin{figure}[htbp]
	\centering
		\includegraphics[width=0.5\textwidth]{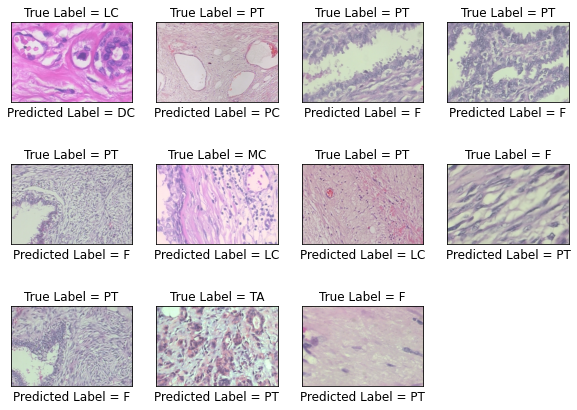}
	\caption{Misclassified samples in the ViT-DeiT model with true and predicted labels, showing that no malignant tumor was classified as a benign tumor. }
	\label{fig:miss}
\end{figure}

The ViT–DeiT model avoids the worst case in cancer diagnosis, that is, the diagnosis of a malignant sample as benign \cite{papageorgiou2018limitations}. The most misclassified images are malignant samples that are predicted as other types of malignant tumors, as shown in Fig. \ref{fig:miss}.

\begin{figure}[htbp]
	\centering
		\includegraphics[width=0.4\textwidth]{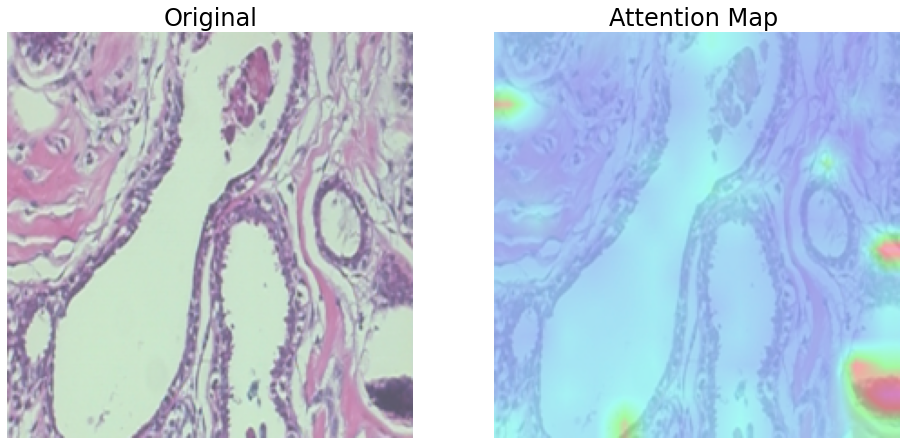}
	\caption{Attention map of the ViT model on sample image from the BreakHis dataset.}
	\label{fig:mapGM}
\end{figure}

\begin{figure}
	\centering
		\includegraphics[width=0.4\textwidth]{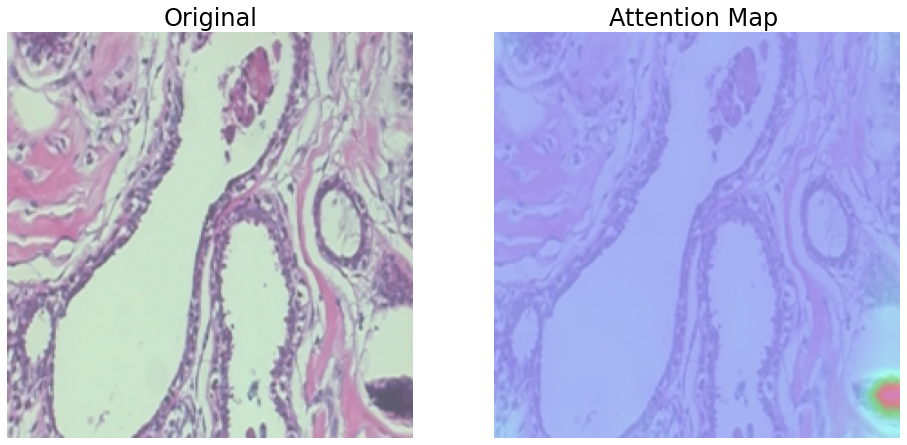}
	\caption{Attention map of the DeiT model on sample image from the BreakHis dataset.}
	\label{fig:mapFM}
\end{figure}

Fig.\ref{fig:mapGM} shows the attention map of the ViT model after training on the BreakHis dataset. Fig.\ref{fig:mapGM} shows that the attention was focused on cancerous cells  and paid little attention to the wrong regions based on expert diagnosis. In addition, the DeiT model accurately paid attention to cancerous cells, as shown in Fig. \ref{fig:mapFM}. The effectiveness of soft voting appears in such cases to minimize the error rate. 


\subsection{Comparison with Similar Works}
 
The performance of the methods in recent studies was compared with that of the proposed ViT–DeiT model. The performance of the multi-class classification with magnification-dependent evaluation is shown in Table \ref{tab:similareW}. Our model achieved the highest results for various magnification factors. Moreover, the approach based on the RANet–ADSVM model \cite{zhou2022breast} achieved an accuracy of 98.05\% at 200× magnification, which is better than in other studies. In addition, the result of our model at 200× magnification factor was close to that of the RANet–ADSVM model, which may be explained by the data balancing step.

\begin{table}[htbp]
\caption{\label{tab:similareW} Comparison of the performance of multi-class classification with magnification-dependent classification with that of similar works }
\centering
\setlength{\tabcolsep}{5pt}
\begin{tabular}{|l|l|l|l|l|l|}
\hline 
Model                                              & \begin{tabular}[c]{@{}l@{}}Magni-\\ fication\end{tabular} & \begin{tabular}[c]{@{}l@{}}Accuracy\\    (\%)\end{tabular} & \begin{tabular}[c]{@{}l@{}}Precision\\    (\%)\end{tabular} & \begin{tabular}[c]{@{}l@{}}Recall\\    (\%)\end{tabular} & \begin{tabular}[c]{@{}l@{}}F1 score\\    (\%)\end{tabular} \\ \hline
\multirow{4}{*}{ Deep-Net \cite{jiang2019breast}}               & 40X           & 94.43                                                      & 95.25                                                       & 95.55                                                    & 95.39                                                      \\
                                                   & 100X          & 94.45                                                      & 94.51                                                       & 94.64                                                    & 94.42                                                      \\
                                                   & 200X          & 92.27                                                      & 90.71                                                       & 92.24                                                    & 91.42                                                      \\
                                                   & 400X          & 91.15                                                      & 90.74                                                       & 91.09                                                    & 90.75                                                      \\\hline
\multirow{4}{*}{ResNet-18 \cite{boumaraf2021new}}           & 40X           & 94.49                                                      & 93.81                                                       & 94.78                                                    & 94.15                                                      \\
                                                   & 100X          & 93.27                                                      & 92.94                                                       & 91.59                                                    & 92.23                                                      \\
                                                   & 200X          & 91.29                                                      & 91.18                                                       & 88.28                                                    & 89.47                                                      \\
                                                   & 400X          & 89.56                                                      & 87.97                                                       & 87.97                                                    & 87.77                                                      \\\hline
\multirow{4}{*}{SE-ResNet \cite{kate2021breast}}                & 40X           & 86.89                                                      & -                                                           & -                                                        & -                                                          \\
                                                   & 100X          & 88.69                                                      & -                                                           & -                                                        & -                                                          \\
                                                   & 200X          & 86.53                                                      & -                                                           & -                                                        & -                                                          \\
                                                   & 400X          & 86.37                                                      & -                                                           & -                                                        & -                                                          \\\hline
\multirow{4}{*}{\begin{tabular}[c]{@{}l@{}}RANet-ADSVM\\\cite{zhou2022breast}\end{tabular}}  & 40X           & 91.14                                                      & -                                                           & -                                                        & -                                                          \\
                                                   & 100X          & 96.83                                                      & -                                                           & -                                                        & -                                                          \\
                                                   & 200X          & 98.05                                                      & -                                                           & -                                                        & -                                                          \\
                                                   & 400X          & 90.30                                                      & -                                                           & -                                                        & -                                                          \\\hline
\multirow{4}{*}{ViT-DeiT (Ours)}           & 40X           & \textbf{99.43}                                             & \textbf{99.38}                                              & \textbf{99.46}                                           & \textbf{99.40}                                             \\
                                                   & 100X          & \textbf{98.34}                                             & \textbf{98.31}                                              & \textbf{98.51}                                           & \textbf{98.35}                                             \\
                                                   & 200X          & \textbf{98.27}                                             & \textbf{98.32}                                              & \textbf{98.27}                                           & \textbf{98.23}                                             \\
                                                   & 400X          & \textbf{98.82}                                             & \textbf{98.57}                                              & \textbf{98.78}                                           & \textbf{98.65}    \\\hline                                        
\end{tabular}
\end{table}

Few studies have focused on magnification-independent fields. Table \ref{tab:MIsimilarW} shows the performance of studies for multiclass classification for the BreakHis dataset with magnification-independent evaluation. Our model achieves the highest results for magnification-independent classification. However, the classification results showed 98.17\% accuracy, 98.18\% precision, 98.08\% recall and 98.12\% F1 score, which were better  than the results (93.32\% accuracy, 92.98\% precision, 92.36\% recall, and 92.44\% F1 score ) obtained with the Xception approach \cite{zaalouk2022deep}.

\begin{table}[htbp]
\caption{\label{tab:MIsimilarW}Comparison of the performance of multi-class classification with magnification-independent classification with that of similar works.}
\centering
\begin{tabular}{|l|l|l|l|l|}
\hline
Model                   & \begin{tabular}[c]{@{}l@{}}Accuracy\\    (\%)\end{tabular} & \begin{tabular}[c]{@{}l@{}}Precision\\    (\%)\end{tabular} & \begin{tabular}[c]{@{}l@{}}Recall\\    (\%)\end{tabular} & \begin{tabular}[c]{@{}l@{}}F1 score\\    (\%)\end{tabular}  \\ \hline
 ResNet-18 \cite{boumaraf2021new}  & 92.03                                                         & 91.39                                                          & 90.28                                                       & 90.77                                                         \\\hline
 6B-Net \cite{jain2020supervised}    & 90.10                                                         & -                                                              & -                                                           & -                                                             \\\hline
  Xception \cite{zaalouk2022deep} & 93.32                                                         & 92.98                                                          & 92.36                                                       & 92.44                                                         \\\hline
ViT-DeiT (Ours)   & \textbf{98.17}                                                & \textbf{98.18}                                                 & \textbf{98.08}                                              & \textbf{98.12}   \\\hline                                            
\end{tabular}
\end{table}

\section{CONCLUSION}
\label{sec:con}
We proposed a classification approach for breast cancer that could classify cancer into eight classes  through an ensemble model to assist pathologists in diagnosis. The BreakHis dataset was used to evaluate the model and to design an ensemble model that integrates the ViT and DeiT models (i.e., ViT–DeiT ensemble model).  The accuracies of the ViT–DeiT model in magnification-dependent multi-class classification were 99.43\%, 98.34\%, 98.27\% and 98.82\% at at 40×, 100×, 200×, and 400× magnification, respectively. In magnification-independent classification, we achieved an accuracy of 98.17\% for multi-class classification. The results of magnification-dependent and -independent classification outperformed state-of-the-art classification methods. Moreover, the ViT– DeiT model avoids misclassifying a malignant tumor as benign. This promising result demonstrates that CAD systems can be trusted to classify breast cancer, which is another step toward digitalization.

\bibliography{References}

\bibliographystyle{ieeetr}

\end{document}